\documentclass[aps,prd,showpacs,preprintnumbers,groupedaddress,twocolumn]{revtex4}
\usepackage{bm} 
\usepackage{latexsym} 
\usepackage{dcolumn}
\usepackage{amsfonts,amssymb} 
\usepackage{graphicx,epsfig}
\usepackage{psfrag} 
\usepackage{amsmath,amssymb}
\usepackage{epsf}

\def\beq{\begin{equation}}
\def\eeq{\end{equation}}
\def\br{\begin{eqnarray}}
\def\er{\end{eqnarray}}
\def\benu{\begin{enumerate}}
\def\eenu{\end{enumerate}}

\begin{document}

\title{Non-Gravitating Scalar Field in the FRW Background}
\author{Nabamita Banerjee}
\email[]{E-mail: nabamita@mri.ernet.in}
\affiliation{Harish-Chandra Research Institute, Chhatnag Road,
Jhunsi, Allahabad 211 019, India.}
\author{Rajeev Kumar Jain}
\email[]{E-mail: rajeev@mri.ernet.in}
\affiliation{Harish-Chandra Research Institute, Chhatnag Road,
Jhunsi, Allahabad 211 019, India.}
\author{Dileep P. Jatkar}
\email[]{E-mail: dileep@mri.ernet.in}
\affiliation{Harish-Chandra Research Institute, Chhatnag Road,
Jhunsi, Allahabad 211 019, India.}
\date{\today}

%%%%%%%%%%%%%%%%%%%%%%%%%%%ABSTRACT%%%%%%%%%%%%%%%%%%%%%%%%%%%%%%%%%%%%%%%%%%

\begin{abstract}
  We study interacting scalar field theory non-minimally coupled to
  gravity in the FRW background.  We show that for a specific choice
  of interaction terms, the energy-momentum tensor of the scalar field $\phi$
  vanishes, and as a result the scalar field does not gravitate.  The
  naive space dependent solution to equations of motion gives rise to
  singular field profile.  We carefully analyze the energy-momentum tensor for
  such a solution and show that the singularity of the solution gives
  a subtle contribution to the energy-momentum tensor.  The space dependent
  solution therefore is not non-gravitating.  Our conclusion is
  applicable to other space-time dependent non-gravitating solutions
  as well.  We study hybrid inflation scenario in this model when
  purely time dependent non-gravitating field is coupled to another
  scalar field $\chi$.
\end{abstract}
\maketitle

%%%%%%%%%%%%%%%%%%%%%%%%%%%%%%%%%%%%%%%%%%%%%%%%%%%%%%%%%%%%%%%%%%%%%%%%%%%%

\section{Introduction}
\label{sec:intro}

Non-minimal coupling of the dilaton and the metric is a generic
feature of any string theory(see e.g., \cite{Polchinski:1998rq}).
While usual compactifications do not generate any potential, in the
low energy limit, for the dilaton or other massless scalars coming
from the string theory, it is generally believed that a potential is
generated for them when one considers flux
compactifications\cite{Polchinski:1998rr, Douglas:2006es}.  Generation
of potential for scalar fields in the flux compactification, makes
them an arena for study of physics of the early universe.  However, if
one looks at models of inflation in these compactifications, a generic
problem crops up in the slow roll inflation.  One of the slow roll
parameters, $\eta$, turns out to be sensitive to the size of
compactified space and one finds that the slow roll condition, i.e.,
$\eta\ll 1$ is harder to achieve\cite{McDonald:2002bd, Riotto:2002yw,
  Holman:2006ek}.  Smallness of the slow roll parameters $\epsilon$
and $\eta$ ensures that the scale factor $a(t)$ in the
Friedmann-Robertson-Walker (FRW) background grows exponentially in
time or alternatively the Hubble parameter $H=a(t)^{-1}da(t)/dt$ is
constant during the inflationary epoch \cite{linde}.

The slow roll conditions come from the Einstein equations, where we
demand that the energy-momentum tensor on the right hand side (RHS) of
the equation is proportional to a constant times the FRW
metric\cite{linde, Bassett:2005xm, Mukhanov:2005sc}.  Since RHS
contains the scalar field energy-momentum tensor this condition
reflects upon the nature of the potential.  However, if we have a
non-minimally coupled scalar field with a potential such that it does
not gravitate, i.e., its energy-momentum tensor vanishes identically
then smallness of slow roll parameters $\epsilon$ and $\eta$ becomes
irrelevant.  This is because throughout the rolling of the
non-gravitating scalar field there is no contribution of it on the RHS
of the Einstein equations.  The shape of the potential and rolling of
the scalar field becomes relevant only when, by some other mechanism,
it starts gravitating.  If we achieve sufficient amount of inflation
before the scalar field starts gravitating then subsequent time
dependence of the scalar field has no bearing on the model of
inflation.  As we will see below, for a specific kind of non-minimal
coupling, one gets a fixed potential for the scalar field so that it
does not gravitate.  It is not obvious if such a non-minimal coupling
and/or the potential can be obtained from a generic flux
compactification.  Nevertheless inflation in the theory with a
non-gravitating scalar field is a novel way of by-passing the
`$\eta$-problem'\cite{McDonald:2002bd} in the models of inflation in
the flux compactifications\cite{Douglas:2006es}.

If we are looking for a non-gravitating scalar field background, it
could generically depend on the spatial coordinates as well.  This
question was addressed by\cite{Ayon-Beato:2005tu} in the Minkowski
spacetime and in the next section we will obtain similar results for
FRW and anti-de Sitter backgrounds.  As in \cite{Ayon-Beato:2005tu,
  demir}, we find that the classical non-gravitating solution allows
space dependent scalar field profile in all these cases, including the
Minkowski spacetime.  However, in every case, this solution is
singular.  That is, the scalar field profile diverges at some point in
space/spacetime.  This singularity begs for a careful evaluation of
the energy-momentum tensor in its neighbourhood.  We analyze the
energy-momentum tensor by regularizing the singular scalar field
profile and find that there is a subtle contribution to the
energy-momentum tensor from the singularity.  This contribution is
proportional to the Dirac $\delta$-function but with an infinite
multiplicative coefficient.  Due to this contribution, the space
dependent scalar field profile actually gravitates.  While this rules
out non-constant solution in the Minkowski spacetime, purely time
dependent solution is allowed class of non-gravitating solutions in
the FRW background.  Similarly, purely radial dependent solution can
be non-gravitating in the AdS background.

Since purely time dependent solution in the FRW background can be
non-gravitating, we build a hybrid inflation model by coupling the
non-gravitating solution to another scalar field $\chi$, which is
gravitating.  Inflation is obtained by rolling of the non-gravitating
scalar field and is exited by rolling of the gravitating scalar field,
{\it i.e.}, $\chi$.  The rolling of $\chi$ is triggered by that of the
non-gravitating field.  Interesting feature of this model is that the
non-gravitating scalar field starts gravitating as soon as $\chi$
starts rolling.

This paper is organized as follows. In section~\ref{sec:ngsfrw}, we
review the result for non-gravitating scalar field and its interaction
potential functional in the Minkowski background and obtain the same
in FRW and AdS backgrounds. In section~\ref{sec:regt}, we perform the
regularized energy-momentum tensor calculation in the vicinity of the
singularity of the scalar field profile. We also mention the
subtleties related to the issue of finding the backreacted metric in
the presence of the non-vanishing stress tensor. In
section~\ref{sec:him}, we propose a novel model of hybrid inflation by
considering another scalar field $\chi$. Finally,
section~\ref{sec:dsc} summarizes our results.

\section{Non-Gravitating Scalar Field in various backgrounds}
\label{sec:ngsfrw}

In this section we will study a scalar field $\phi({\bf x}, t)$
non-minimally coupled to gravity.  We will first recall how the
non-gravitating solution is obtained in the Minkowski
space\cite{Ayon-Beato:2005tu}.  We will then obtain the solution in
the FRW background as well as in the anti-de Sitter space.
We are looking for a solution to the
equations of motion such that the energy-momentum tensor for the field
$\phi({\bf x}, t)$ evaluated on this background vanishes.  It imposes
constraints which are sufficient to determine the form of the
classical solution as well as the potential $V(\phi)$.  The equation
of motion of $\phi({\bf x}, t)$ then becomes a consistency condition
for the solution and the potential energy functional.

Let us consider a self-interacting scalar field $\phi({\bf x}, t)$
non-minimally coupled to gravity with cosmological constant $\Lambda$ 
in $(3+1)$ dimensions.
\begin{eqnarray}
  \label{eq:Action}
 I_{\phi}&=&\int{\mathrm{d}^{4}x}\sqrt{-g}\Biggl( \frac{1}{2\kappa
  }(R+2\Lambda) \nonumber \\
 &&  +\frac{1}{2}\phi \square \phi -\frac{1}{2}\xi R\,\phi
  ^{2}-V(\phi )\Biggr). 
\end{eqnarray}
The parameter $\xi$ is a measure of non-minimality.  In 3+1
dimensions, $\xi=1/6$ leads to conformal coupling of the scalar
field to the gravity and $\xi=0$ corresponds to the usual minimal
coupling.  The field equation is
\begin{equation}
  \label{eq:KG}
(\square -\xi R) \phi = V'(\phi )\;,
\end{equation}
where the additional term proportional to the Ricci scalar is a
consequence of the non-minimal coupling.  The conserved
energy-momentum tensor is also modified due to the non-minimal
coupling and is given by
\begin{equation}
  \label{eq:improvement}
\Theta_{\mu \nu }\equiv T_{\mu \nu }+\xi (g _{\mu \nu} \square
-\nabla _{\mu }\nabla _{\nu }+\Lambda\,g_{\mu \nu})\phi ^{2} ,  
\end{equation}
where, $T_{\mu \nu}$ is the standard energy-momentum tensor for a minimally
coupled scalar field,
\begin{equation}
  \label{eq:Tmn}
T_{\mu \nu }=\partial_{\mu}\phi \partial_{\nu }\phi - g_{\mu \nu }\biggl(
  \frac{1}{2}\partial _{\alpha}\phi \partial ^{\alpha }\phi +V(\phi )\biggr).  
\end{equation}
The expression for the energy-momentum tensor in
eq.(\ref{eq:improvement}) is obtained in the following way.  We first
assume that in its original form, every component of the
energy-momentum tensor $\Theta_{\mu\nu}$ vanishes.  In that case the
Einstein equation becomes $G_{\mu\nu} = \Lambda g_{\mu\nu}$.  Using
this we arrive at the form of $\Theta_{\mu\nu}$ given in
eq.(\ref{eq:improvement}).

So far we have been looking at the general features of a scalar field
non-minimally coupled to the gravity.  
Our aim is to look for a scalar field solution which does not
gravitate in a given background.  The solution is obtained by first doing
a change of variable
\begin{equation}
  \label{eq:sigma}
\phi = \sigma^{\alpha}\; , 
\end{equation}
in the expression for the energy-momentum tensor (\ref{eq:improvement}) and
writing them 
in terms of $\sigma$ and its derivatives.  Setting every component of
$\Theta_{\mu\nu}$ to zero gives us a set of equations.
\begin{subequations}
\begin{eqnarray}
\label{eq:Tmunu} 0\!&\!=\!&\!\Theta
_{\mu\nu}=\frac{(2\xi)^2}{(1-4\xi)}\frac{\phi^2}{\sigma}
\nabla_{\mu}\partial_{\nu}\sigma, \qquad \mu\neq\nu,\\
\label{eq:Tii}
0\!&\!=\!&\!g_{ii}\Theta_{tt}-g_{tt}\Theta_{ii}\nonumber \\
&\!=\!&\frac{(2\xi)^2}{(1-4\xi)}
\frac{\phi^2}{\sigma}
\left(g_{ii}\nabla_t\partial_{t}\sigma-g_{tt}\nabla_i\partial_i\sigma\right) 
\;\mbox{(no sum)},\\
\label{eq:U}
0\!&\!=\!&\!\Theta_{tt}\nonumber \\
&\!=\!& V(\phi)\!-\!\frac{(2\xi)^2\phi^2}{(1-4\xi)\sigma}\!
\left[\!\frac{\partial_\rho\sigma\partial^\rho\sigma}{2(1-4\xi)\sigma}
  \!-\!\!g^{ii}\sum_{i=1}^{3}\nabla_i\partial_i\sigma\right]\nonumber \\
&& - \xi\phi^2\Lambda .
\end{eqnarray}
\end{subequations}
These equations impose constraints on $\sigma$.  These equations are
obtained by choosing $\alpha= 2\xi/(4\xi -1)$, where $\xi \neq 0,
1/4$.  While the case $\xi=0$ is excluded because it takes us back to
the minimally coupled scalar field theory, $\xi=1/4$, can be treated
separately and it can be shown that non-gravitating solution exists
for $\xi=1/4$ as well\cite{Ayon-Beato:2005tu}. We will solve these
constraints in various backgrounds below.

{\em The Minkowski Space}: Let us start with the Minkowski space
background.  The equations (\ref{eq:Tmunu}) and (\ref{eq:Tii}) are
solved in the Minkowski space by\cite{Ayon-Beato:2005tu}
\begin{equation}
  \label{eq:flatsig}
  \sigma(x^{\mu}) = a_1 x^{\mu}x_{\mu} + p^{\mu}x_{\mu} + a_2,
\end{equation}
and the self interacting potential is given by
\begin{equation}
  \label{eq:potmin}
 V(\phi) = \frac{2\xi^2}{(1-4\xi)^2}(\lambda\phi^{(1-2\xi)/\xi} +
  48 (\xi-\frac{1}{6}) a_1 \phi^{1/2\xi}),
\end{equation}
where, $\lambda = p^{\mu}p_{\mu} - 4a_1a_2$.  We will get back to
this solution in the next section.  

{\em The FRW Space}: Let us now turn our attention to the FRW space.
Background metric in this case is given by
\begin{equation}
  \label{eq:met}
  ds^2 = -dt^2+a^2(t)(dx^2+dy^2+dz^2)
\end{equation}
where $a(t)$ is the scale factor of the spatial section of the
background.  By using the change of variables
(\ref{eq:sigma}) and rearranging components of the energy-momentum tensor
(\ref{eq:improvement}) appropriately we get three independent set of
expressions, all of which are set to zero.

\begin{figure*}[t]
  \centering {\includegraphics[scale=0.8]{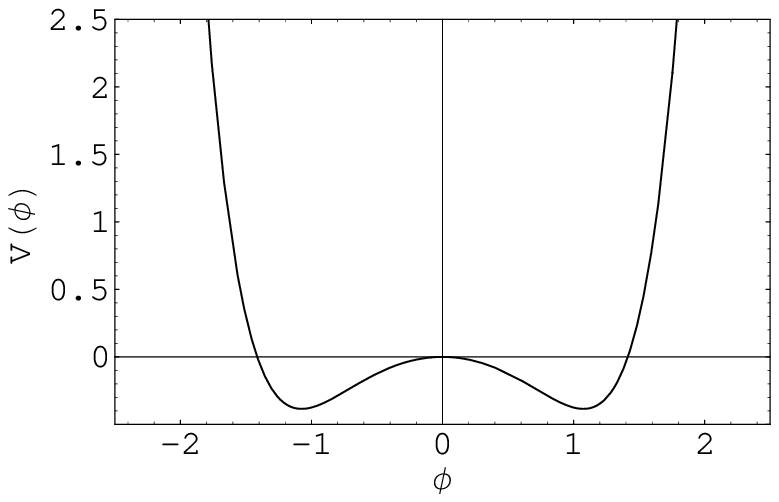}(a)
    \includegraphics[scale=0.8]{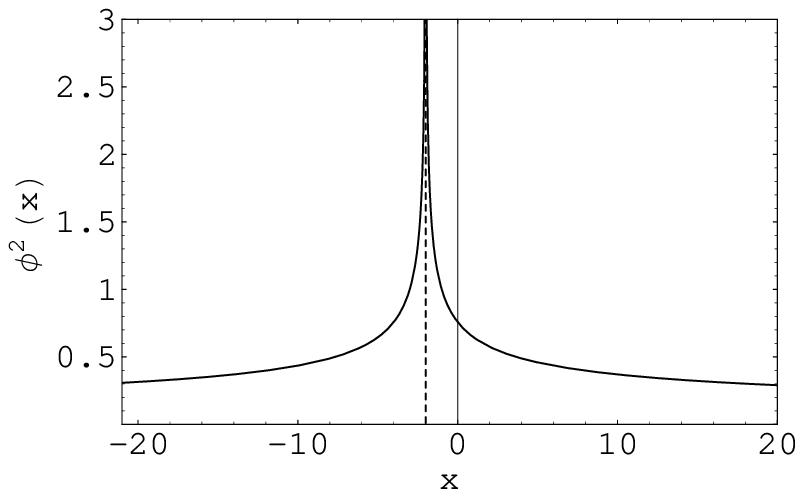}(b)}
  \caption{\label{fig:phi} Fig.(a) shows generic form of the potential
    for the non-gravitating field $\phi$ in the FRW background.  Curve
    in fig.(b) denotes $\phi^2$, square of the classical solution at a
    fixed time.}
\end{figure*}

Eq.~(\ref{eq:Tmunu}) imposes constraints on the profile of $\sigma$.
When neither $\mu$ nor $\nu$ is equal to $t$ then the constraints on
$\sigma$ are such that we choose
\begin{equation}
  \label{eq:ansatz}
\sigma =f(t)(X_1(x)+X_2(y)+X_3(z)).  
\end{equation}
If instead either $\mu$ or $\nu$ is equal to $t$ then the constraint
coming from the equation (\ref{eq:Tmunu}) implies the function $f(t)$
is proportional to the scale factor $a(t)$, i.e.,
\begin{equation}
  \label{eq:ftat}
  f(t) = c_0 a(t).
\end{equation}
Linear combination of the diagonal components of the energy-momentum
tensor in the eq.(\ref{eq:Tii}) determines the solution completely.
That is, while on one hand, it can be used to obtain behaviour of the
scale factor, on the other hand, it also determines the functional
forms of $X_1(x)$, $X_2(y)$ and $X_3(z)$ up to constants of
integration.  This equation also requires exponential behaviour of the
scale factor.  This is consistent with the Einstein equation which can
be written in terms of the Hubble parameter $H= a(t)^{-1}(da(t)/dt)$
and the cosmological constant $\Lambda$ as,
\begin{equation}
  \label{eq:cchub}
  3H^2 = -\Lambda, \qquad a(t) = a_0\exp(Ht).
\end{equation}
Vanishing of the double spatial derivative of $\sigma$, in
eq.(\ref{eq:Tii}), implies that the functions $X_i$ are at most linear
in their arguments for all $i$, for example,
\begin{equation}
  \label{eq:exone}
  X_1(x) = p_1 x + c_1.
\end{equation}
Using eqs.(\ref{eq:ansatz}--\ref{eq:exone}) in the relation
(\ref{eq:sigma}) we get
\begin{equation}
  \label{eq:phfrsm}
  \phi({\bf x}, t) = \sigma^{2\xi/(4\xi-1)} = \left(c_0 a(t)
    (p_ix^i + c) \right)^{2\xi/(4\xi-1)},
\end{equation}
where, $c= c_1+c_2+c_3$ and $p_ix^i = p_1x+p_2y+p_3z$.  We can now
determine the potential $V(\phi)$ using the $\Theta_{tt}$ equation and
the solution (\ref{eq:phfrsm}).
\begin{eqnarray}
  \label{eq:vphi}
  V(\phi) &=&
  \left[\frac{2\xi^2}{(4\xi-1)^2}(5-24\xi)-3\xi\right]
  H^2\phi^2\nonumber \\
 && + \frac{2\xi^2}{(4\xi-1)^2}
 \left( c_0^2\sum_{i=1}^3 p_i^2\right)\phi^{(1-2\xi)/\xi}.
\end{eqnarray}
Let us first notice that for all $\xi<1/4$, second term in the
potential is dominant for large $\phi$ and if $\xi$ is such that the
potential $V(\phi)$ is even under $\phi\to -\phi$ then the potential
is bounded from below.  Since the first term is negative, for the
parity symmetric cases we have a double well
potential(fig.\ref{fig:phi}(a)).  The solution (\ref{eq:phfrsm})
obtained by solving the energy-momentum tensor constraints also
satisfies the $\phi$ equation of motion.  This consistency condition
ensures that the energy-momentum tensor is conserved in this
background.
This solution diverges for large $x^i$ when $\xi>1/4$($\alpha>0$), on
the other hand, for $\xi<1/4$($\alpha<0$) it diverges at a point where
$(p_ix^i + c)=0$(fig.\ref{fig:phi}(b)).  
We choose $\xi$ in such a way that $\alpha$, which in
general is a fraction, has odd denominator.  This ensures that at
least one real solution always exists.  The classical solution
$\phi({\bf x}, t)$ depends on the scale factor $a(t)$ and when
$\xi<1/4$, the field $\phi({\bf x}, t)$ is inversely proportional to a
$\xi$ dependent power of $a(t)$.  Thus, in addition to having the
space dependent profile, the amplitude of the classical solution
depends on the scale factor and decreases rapidly for $\xi<1/4$.

For the sake of completeness let us consider the case of $\xi=1/4$.
In this case we carry out analysis similar to that of
\cite{Ayon-Beato:2005tu} and obtain the classical solution $\phi({\bf
  x}, t)$ and the potential $V(\phi)$ exactly.
\begin{equation}
  \label{eq:quarpro}
   \phi({\bf x}, t) = \exp(c_0 e^{Ht}(p_ix^i+c)),
\end{equation}
\begin{eqnarray}
\label{eq:quarzet}
 V(\phi)&=&\phi^2\Biggl[\sum_{i=1}^3 
  \frac{(c_0p_i)^2}{2}
  -\frac{3H^2}{4} -\frac{(H\ln\phi)^2}{2}\nonumber\\
 && -\frac{3H^2}{2}\ln\phi \Biggr].
\end{eqnarray}

{\em The Anti-de Sitter space}: Let us now turn our attention to the
Anti-de Sitter space(AdS).  The metric on the four dimensional AdS
space is given as,
\begin{equation}
ds^2= -\frac{r^2}{b^2} dt^2 +\frac{b^2}{r^2} dr^2 + \frac{r^2}{b^2}
(dx^2 + dy^2),
\end{equation}
where, $b$ is the radius of the anti-de Sitter space.  The conditions
coming from the $\Theta_{\mu\nu}$ ($\mu\neq \nu$) can be satisfied by
choosing the following ansatz for $\sigma$
\begin{equation}
\sigma= a(r)[f_1(t)+f_2(x)+f_3(y)].
\end{equation}
Diagonal components of the energy-momentum tensor then give
\begin{equation}
a(r) = C r,
\end{equation}
where C is the integration constant and
\begin{equation}
\sigma= C r (t+x+y+d),
\end{equation}
where $d$ is another constant of integration.  The potential is
obtained by solving $\Theta_{tt} = 0$.
\begin{equation}
  V(\phi)=\alpha^2 b ^2 C^2 \phi^{\frac{(1- 2 \xi)}{\xi}} +
  (8\xi \alpha + 5\xi + \frac{\alpha^2}{2}) \frac{\phi^2}{b^2}.
\end{equation}
It is worth emphasizing that this solution solves the equation of
motion ensuring conservation of the energy-momentum tensor.

While we have managed to obtain non-gravitating solutions to the
equations of motion in various backgrounds, all these solutions are
singular. If one chooses $\xi > 1/4$ then the solution diverges at
infinity and if $\xi < 1/4$ then it diverges for some finite value of
space-time coordinates.  This divergence in the solution has a subtle
contribution to the energy-momentum tensor.  In the next section we
will carefully analyze the solution and show that naive conclusion of
vanishing of the energy-momentum tensor is flawed.

\section{Regularized energy-momentum tensor}
\label{sec:regt}
In this section we will show that the naive computation of the stress
tensor for the ``non-gravitating'' solution in the previous section
fails in the vicinity of the singularity.  Singular nature of the
solution, depending on the choice of $\xi$, leads to either
discontinuous profile for $\phi$ or its derivative
$\partial_{\mu}\phi$.  Since the energy-momentum tensor contains terms
upto second derivative of $\phi$, this discontinuity can lead to a
potential $\delta$-function contribution.  To analyze this we will
regularize the solution in the neighbourhood of singularity.  We will
illustrate this method for the Minkowski space as well as for the FRW
space, a similar analysis carries through for the anti-de Sitter
space.  For concreteness, we will carry out our analysis for
$\xi<1/4$, but it can be easily extended to $\xi>1/4$.  While in the
former case the singularity is in the finite domain, it is at infinity
for the latter case.

{\em The Minkowski Space}: In this case we regularize solution by
regularizing the function $\sigma$ as follows
\begin{equation}
\sigma = \sqrt{(x+t)^2+\epsilon ^2} \sqrt{(x-t)^2+\epsilon ^2},
\end{equation}
where, for illustration, we have chosen $a_1=1$, $p_{\mu} =0$ and
$a_2=0$ in (\ref{eq:flatsig}).  It is evident that as $\epsilon
\rightarrow 0$, we recover original form for $\sigma$, namely $\sigma
= x^2-t^2$.  Notice that we are choosing the profile for $\sigma$ such
that it depends only on one spatial dimension.  This suffices to
illustrate our point and reinstating its full space dependence does
not alter our conclusion.  Using this regularized form of $\sigma$,
and therefore $\phi$, we can write down any component of the
energy-momentum tensor $\Theta_{\mu\nu}$ at an arbitrary point in
Minkowski space.  The non-vanishing off-diagonal component of
$\Theta_{\mu\nu}$ is,
\begin{equation}
\label{eq:ttx}
\Theta_{tx}=-2\xi\alpha
\sigma^{2\alpha}\left[\frac{\epsilon^2}{((x+t)^2+\epsilon ^2)^2}-
\frac{\epsilon^2}{((x-t)^2+\epsilon ^2)^2}\right].
\end{equation}
At any point where the field profile $\phi$ is finite, we can show
using independent methods that every component of the energy-momentum
tensor vanishes.  We, therefore, need to restrict ourselves to the
neighbourhood of the singularity.  Taking $(x+t)\sim \epsilon$ and
$(x-t)\sim \epsilon$,
\begin{equation}
\label{eq:ttxreg}
\Theta_{tx}=-2\xi\alpha \epsilon^{4\alpha}
f_1\left(\frac{x}{\epsilon},\frac{t}{\epsilon},\epsilon\right).
\end{equation}
As expected the function $f_1$ is such that when $x,t \gg \epsilon$ it
vanishes as $\epsilon\rightarrow 0$, and when $x,t \sim \epsilon$,
$f_1 \sim \epsilon^{-2}$.  Therefore, in the neighbourhood of the
singular point, {\em i.e.}, when $x,t\sim \epsilon$
\begin{equation}
  \label{eq:xtdiv}
  \Theta _{tx} \sim \epsilon^{(4\alpha-2)}.
\end{equation}
Since $4\alpha-2$ is negative for $\xi<1/4$, we see that the
$\Theta_{tx}$ component of the energy-momentum tensor is divergent in
the limit $\epsilon\to 0$.  In fact, analysis of diagonal components
of the energy-momentum tensor shows that they also diverge as
$\epsilon^{(4\alpha-2)}$ in the neighbourhood of the singular point.
Notice that in the limit $\epsilon\to 0$ the function $f_1$ has
$\delta$-function support at the location of the singularity, however,
the coefficient multiplying this $\delta$-function diverges as we
remove the regulator.  This severe singular contribution to
$\Theta_{\mu\nu}$ invalidates the naive derivation of the
non-gravitating solution in the previous section.

{\em The FRW Space}: The situation is not any better in the FRW
background. We write the regularized solution as
\begin{equation}
\label{eq:phfrw}
\phi = a(t)^{\alpha} x (x^2+\epsilon^2)^\beta.
\end{equation}
where $\beta=(\alpha-1)/2.$ Substituting this regularized form of the
solution in the energy-momentum tensor, we find that all off diagonal
components of $\Theta_{\mu\nu}$ vanish identically at the singular
point.  The diagonal components, on the other hand, take the form
\begin{equation}
\label{eq:ttfrw}
\Theta_{tt}=  a(t)^{2\alpha}\epsilon^{4\beta}
f_2\left(\frac{x}{\epsilon}\right).
\end{equation}
The function $f_2$ vanishes when $ x \gg \epsilon$ and $\epsilon
\rightarrow 0$ and when $x \sim \epsilon$, it is constant.
Therefore near the singular point {\em i.e.} when $x \sim \epsilon$ and 
$\epsilon \rightarrow 0$, $\Theta _{tt}$ diverges as $\epsilon^{4\beta}$. 
All other diagonal components exhibit the similar divergent behaviour,
\begin{equation}
\label{eq:iifrw}
\Theta_{ii}=  a(t)^{2\alpha+2}\epsilon^{4\beta-2}
f_2\left(\frac{x}{\epsilon}\right).
\end{equation}
This divergence is as severe as in the case of the Minkowski space.
Analysis of the energy-momentum tensor of the scalar field in the
anti-de Sitter space proceeds in the similar fashion as in the FRW
case.

We can now ask what is the back reaction of this contribution to the
energy-momentum tensor on the background metric.  Notice that in the
FRW case the Dirac $\delta$-function contribution exists only for the
diagonal components of the energy-momentum tensor and all the
off-diagonal components continue to vanish.  With this in mind we
consider the metric ansatz for the backreacted geometry to be
\begin{equation}
\label{eq:brmet}
g_{\mu\nu}= diag(-f_1(x,t), f_2(x,t), f_3(x,t), f_3(x,t)).
\end{equation}
Substituting this ansatz in the Einstein equation gives us a set of
coupled non-linear inhomogeneous partial differential equations.  To
make things even harder, we find, coefficient of the inhomogeneous
term is time dependent.  In order to seek a solution we will attempt a
separable ansatz for the solution.  We will also look at the solution
on one side of the singularity.  A similar solution on the other side
of the singularity can then be matched with the first one along with
the contribution from the singularity.  Since the energy-momentum
tensor vanishes away from the singularity, RHS of the Einstein
equation is proportional to the cosmological constant term only.  As
an illustration let us consider an ansatz
\begin{equation}
\label{eq:sepan}
f_i(x,t) = (cx+d)^{\alpha_i}g_i(t).
\end{equation}
With this ansatz, we get one sided solution with $\alpha_1=\alpha_3=0$
and $\alpha_2 = -2$ and two classes of time dependent terms are
$g_1(t)=c_1$, $g_2(t)=c_2e^{-2t}$, $g_3(t)=c_3e^{t}$ or $g_1(t)=c_1$,
$g_2(t)=c_2e^{2t}$, $g_3(t)=c_3e^{-t}$ where $c_1$, $c_2$ and $c_3$
are integration constants. We can now take the $\delta$-function
contribution into account to match solutions on either side of the
singularity.  Due to the time dependent coefficient of the
$\delta$-function, we end up with inconsistent powers of the scale
factor on the right and left hand side of the Einstein equation.  We
therefore conclude that the separable ansatz does not give a solution
to our coupled inhomogeneous PDEs.  We have tried out more general
separable ansatze but all of them have suffered the same fate.

Since vanishing of the energy-momentum tensor also gives rise to
coupled nonlinear, but homogeneous, partial differential equations, we
can attempt to get another solution of these equations which are not
singular.  This does not seem to lead to any new solution.  Noticing
the fact that the scalar field profile is proportional to $\exp(Ht)$,
we can try a field redefinition, $\phi = \exp(\psi)$ and attempt to
seek a solution
\begin{equation}
\label{eq:frexpo}
\psi(x,t) = At + f(x).
\end{equation}
This ansatz gives us back the singular solution.  Instead of the
additive ansatz if we take a multiplicative ansatz, i.e., $\psi(x,t) =
Atf(x)$ we end up with a contradiction.  While one equation demands
the scale factor to be a function of time only, other equation is
satisfied only if the scale factor has spatial dependence.  Replacing
$At$ in the product ansatz by any other time dependent function does
not alter the result.

\section{Hybrid Inflation Model}
\label{sec:him}

In the previous section, we saw that space dependent scalar field
profile fails to be non-gravitating. However, if we look at purely
time dependent scalar field profile in the FRW background, it does not
lead to any subtle contribution to the energy-momentum tensor.  We can
therefore use the space independent scalar field profile to construct
a model of inflation.

To do that we will couple another scalar field $\chi({\bf x}, t)$ to
the model studied above.  For simplicity we consider minimal coupling
between $\chi({\bf x}, t)$ and gravity.  The additional term in the
action corresponding to the field $\chi({\bf x}, t)$ is
\begin{equation}
  \label{eq:Lch}
  I_{\chi} = \int{\mathrm{d}^{4}x}\sqrt{-g}\left(\frac{1}{2}\chi
    \square \chi -V_{\chi}(\phi, \chi)\right). 
\end{equation}
We choose the interaction term $V_{\chi}$ as
\begin{equation}
  \label{eq:vcin}
  V_{\chi}(\phi, \chi)
  =\frac{\lambda_1}{2}\left(\phi^2-\phi_0^2\right)\chi^2 +
  \lambda_2 \chi^4,
\end{equation}
where, $\lambda_i$ are positive for $i=1,2$.  We define the total
action to be $I= I_{\phi}+ I_{\chi}$.  We can now justify the
cosmological constant term in $I_{\phi}$ by interpreting it as the
value of the potential $V_{\chi}$ at $\chi=0$.  This essentially
amounts to redefining value of $V_{\chi}$ at $\chi=0$.  For the model
to be studied here, we will choose $\xi<1/4$ therefore, in this case
magnitude of the classical solution $\phi({\bf x}, t)$ decreases as a
function of time as the scale factor $a(t)$ grows.

For a fixed $\phi_0$, it is easy to see from the potential
(\ref{eq:vcin}), that we have two regimes corresponding to $\phi >
\phi_0$ and $\phi < \phi_0$.  As we discussed in the previous section,
space dependent profile for $\phi$ is singular and it gives rise to a
subtle contribution to the energy-momentum tensor.  The space dependent
solution (\ref{eq:phfrsm}) therefore is not a bona fide non-gravitating
solution.  Purely time dependent configuration for $\phi$, however,
does not suffer from this problem and is a legitimate non-gravitating
field configuration.  For $\phi > \phi_0$, the potential for the field
$\chi$ has a unique minimum at $\chi =0$ and in this case the field
$\phi$ does not gravitate.  The exponential growth of scale factor is
then determined solely by the cosmological constant $\Lambda = \langle
V_{\chi}\rangle|_{\chi=0}$ and the rate at which the field $\phi$ is
rolling is irrelevant.  In particular, the slow roll parameters
$\epsilon$ and $\eta$ determined from the potential $V(\phi)$ have no
bearing on the nature of the inflation as $\phi({\bf x}, t)$ is not
gravitating in this domain.  This situation continues until $\phi({\bf
  x}, t_0)= \phi_0$ at some time $t=t_0$.  For all subsequent times
after $t_0$, we are in a situation where $\phi < \phi_0$.  A
sufficient amount of inflation can be obtained by adjusting initial
value of $\phi$.

Once $\phi < \phi_0$, the field $\chi$ develops tachyonic instability
at $\chi=0$.  This instability makes $\chi$ roll towards the true
minimum.  The rolling of $\chi$ affects exponential growth of the
scale factor.  This is because two new terms are generated due to
rolling of the field $\chi$.  First of all, rolling of $\chi$ brings
in $I_{\chi}$ into the picture, {\em i.e.}, the field $\chi$
contributes to the energy-momentum tensor and consequently to the
right hand side of the Einstein equation. Secondly, non-zero vacuum
expectation value of $\chi$, changes the potential $V(\phi)$ by adding
a $\phi^2$ term which modifies mass of $\phi$.  Due to this additional
term the field $\phi$ ceases to be a non-gravitating field and starts
contributing to the right hand side of the Einstein equation.  The
first term is identical to the scalar field in the hybrid inflation
models, which is responsible for exit from the inflationary regime.
It would be interesting to study this model further and check its
validity against currently available CMB data\cite{Spergel:2006hy,
  Jeannerot:2005mc}.

\subsection{Conformal Transformation}
\label{sec:ct}

It is known for some time that action of the scalar field nonminimally
coupled to gravity can be brought into canonical minimally coupled
form by conformal transformation(see {\it e.g.},
\cite{Faraoni:1998qx}).  Consider the action
\begin{equation}
\label{eq:nmin}
S = \int d^4x \sqrt{-g} \left[\frac{\Omega^2}{2\kappa}R - 
\frac{1}{2}g^{\mu\nu}\partial_{\mu}\phi\partial_{\nu}\phi 
- V(\phi) \right],
\end{equation}
where, $\Omega^2 = 1-\kappa\xi\phi^2$.  A minimally coupled scalar
field theory can be obtained by doing the conformal transformation,
\begin{equation}
\label{eq:cftr}
\Omega^2 g_{\mu\nu}= \tilde g_{\mu\nu},
\end{equation}
and followed by the redefinition of the scalar field
\begin{equation}
\label{eq:redsc}
d\tilde\phi = \frac{[1-\kappa\xi(1-6\xi)\phi^2]^{1/2}}{\Omega^2}d\phi
\, .
\end{equation}
The new action written in terms of $\tilde\phi$ with the potential energy
functional
\begin{equation}
\label{eq:tilpot}
\tilde V(\tilde\phi) = \frac{V(\phi)}{\Omega^4},
\end{equation}
is minimally coupled to gravity.  The equation (\ref{eq:redsc}) can be 
integrated to write $\tilde\phi$ as a function of $\phi$
\begin{eqnarray}
\label{eq:ptilp}
\tilde\phi &=& \frac{1}{\sqrt{2\kappa}\xi}\Biggl( 2\sqrt{3}\xi\tanh^{-1}
\biggl(\frac{\sqrt{6\kappa}\xi\phi}
{\sqrt{1-\kappa\xi(1-6\xi)\phi^2}}\biggr)\nonumber \\[2mm]
&&\!\!\!\!\!  -\sqrt{2\xi(-1+6\xi)}\sinh^{-1}(\sqrt{\kappa\xi(-1+6\xi)}\phi)
\Biggr) .
\end{eqnarray}
Substituting this redefinition in the action gives rise to a pretty
complicated potential energy functional.  Nevertheless, in principle,
it is possible to get non-gravitating scalar field background even in
the minimally coupled case.  This result holds for all backgrounds
including the Minkowski space\cite{Ayon-Beato:2005tu}.  The
non-minimally coupled situation is, clearly, much easier to deal
with.

In case of the hybrid inflation model, we have an added complication
in the minimally coupled situation.  This is due to the fact that the
conformal transformation which relates the non-minimally coupled
theory to the minimally coupled one shows up in front of the kinetic
energy term of $\chi$ as well as the potential energy term.  The
action for $\chi$ becomes
\begin{equation}
  \label{eq:actchi}
\mathcal{L}_{\chi} = - \Omega^{-2}\tilde g^{\mu\nu}
\partial_{\mu}\chi\partial_{\nu}\chi - \Omega^{-4} V_{\chi}(\tilde\phi, \chi).
\end{equation}
Since $\Omega$ is field dependent, it generates a non-trivial metric
in the field space.

\section{Discussion and Summary}
\label{sec:dsc}

We have considered a novel model for inflation with a non-minimally
coupled non-gravitating scalar field.  Naive non-gravitating classical
solution for the scalar field $\phi$ has space dependent profile.
Such a space dependent profile is singular.  This singularity gives
rise to a subtle but divergent contribution to the energy-momentum
tensor.  Due to this contribution, the space dependent profile of
$\phi$ fails to be a non-gravitating solution.  The divergent
contribution to the energy-momentum tensor has a delta function
support at the location of singularity of the classical solution.  The
divergence is more severe than just the delta function because for any
$\alpha\neq 0$ the coefficient multiplying the delta function also
diverges.  This problem is not specific only to the space dependent
profile of the scalar field in the FRW background.  This problem
exists for solutions in the Minkowski space and the anti-de Sitter
space as well.  We attempted to determine backreaction of this
divergent contribution on the geometry.  This gives rise to a set of
coupled inhomogeneous non-linear partial differential equations.
Several obvious and not-so-obvious ansatz fail to solve these
equations.

Space independent profile for $\phi$, however, does retain the
non-gravitating property.  We use this fact to write down a hybrid
inflation model by coupling $\phi$ to another scalar field $\chi$.
Utility of this model is in the non-gravitating nature of the scalar
field $\phi$.  Due to this property the inflation is completely driven
by the cosmological constant, which is the value of the potential
$V_{\chi}(\phi, \chi)$ at $\chi=0$. Expansion of the universe is
independent of how $\phi$ rolls down and enters into the gravitating
regime.  Value of slow roll parameters $\epsilon$ and $\eta$ is
therefore of little consequence.  Although non-minimally coupled
scalar field can, after a conformal transformation, be brought into
minimally coupled form, we show that such a field redefinition is not
very convenient.  It is, in fact, more efficient to work with
non-minimal coupling.

\vspace*{15pt}

We would like to thank J. S. Bagla, S. Dutta, R.  Gopakumar, T.
Padmanabhan and L. Sriramkumar for numerous discussions.  We would
especially like to thank A. Sen for many an interesting discussions
and for emphasising on the regularized energy-momentum tensor near
the singularity.  We would also like to thank D. Choudhury for useful
correspondence and pointing out subtleties involved in seeking
numerical solutions to coupled inhomogeneous PDEs.

\end{document}